\documentclass[12pt]{article}
\usepackage{amsmath}
\usepackage{epsfig}
\usepackage{latexsym}
\usepackage{a4wide}
\usepackage{bbm}
\usepackage{feynmf}
\newcommand{\I}{\text{i}}
\newcommand{\E}{\text{e}}
\newcommand{\tr}{\text{tr}}

\newcommand{\re}[1]{(\ref{#1})}
\newcommand{\sta}[1]{{}^\star\! #1}
\begin{document}
\abovedisplayskip17pt plus2pt minus4pt
\abovedisplayshortskip14pt plus2pt minus4pt
\belowdisplayskip17pt plus2pt minus4pt
\belowdisplayshortskip14pt plus2pt minus4pt
\title{Second-Order Radiative Corrections to the Axial Vector Anomaly}
\author{Walter Dittrich\\
  Institut f\"ur theoretische Physik, Universit\"at T\"ubingen,\\
  72076 T\"ubingen, Germany}
\maketitle
\begin{abstract}
We re-examine the historically important decay of the neutral pion
into two photons. Schwinger's Equivalence Theorem is confirmed. We
then consider radiative corrections to the famous Adler-Bell-Jackiw
(ABJ) anomaly. The result depends crucially on a physically motivated
regularization scheme. Our approach is largely based on Schwinger's
source (dispersion) method.
\end{abstract}

\section{Introduction}
From time to time people get excited about the question as to whether
the classic one-loop triangle ABJ anomaly \cite{1} obtains
higher-order loop corrections. If so, what are the consequences of
this modified anomaly for the $\pi^0 \to 2 \gamma$ decay mode which is
stated to one-loop order in form of a local action term \cite{2} 
\begin{displaymath}
W_{\pi^0 \to 2\gamma} =\frac{g}{m} \left( \frac{\alpha}{\pi} \right)
\int d^4x\, \phi(x)\, \mathbf{E\cdot B}(x). 
\end{displaymath}
It is almost thirty years ago that Lester L. DeRaad, Kim Milton, and
Wu-yang Tsai \cite{3} challenged the prevalent view that the ABJ
triangle anomaly is an exact statement. That no further corrections
are expected is claimed by the so-called Adler-Bardeen theorem
\cite{4}. Adler and Bardeen considered ultraviolet regularization and
showed that three-particle exchange processes are not divergent and
therefore do not contribute to the anomaly. But there is also the
two-particle intermediate state with the pseudoscalar form factor to
be taken into account. This contribution cannot be regularized in a
chiral invariant way and so allows for a certain freedom in choosing a
normalization point. Hence the occurrence of an anomaly correction
depends crucially on the way the infrared regularization is
performed. The anomaly as a short-distance or high-energy phenomenon
with cutoff $\Lambda$ has to be supplemented with the lower end of the
momentum scale, $\mu$. Source theoretical calculations show explicitly
that it is indeed not sufficient to merely consider an ultraviolet
regularization but an infrared regularization as well. Incidentally,
this fact was known a long time ago to the late J. Schwinger who
showed in the last chapter of his monograph, Ref. \cite{5}, how the
Equivalence Theorem on the next dynamical level becomes modified by
the replacement $\frac{\alpha}{\pi} \to \frac{\alpha}{\pi} \left( 1+
  \frac{\alpha}{2\pi} \right)$ in the original one-loop triangle
diagram. 

More recently corrections to the one-loop chiral anomaly were also
discussed by V.I. Zakharov \cite{6} and by the authors in
Ref. \cite{7}. The greatest impetus on the matter came, however, from
a new approach via the so-called average-effective action which
M. Reuter published in Ref. \cite{8}. Although we do not have the time
to enter this subject, the reader is invited to consult that paper
(especially the appendix) for further details. 

\section{The Equivalence Theorem, Prehistory}

In his seminal work, Schwinger \cite{2} proved the ``Equivalence
Theorem'' which states that in the low-energy regime, a pseudoscalar
interaction between a spinless neutral meson and a fermion leads to
the same result for the meson decay into two photons as a pseudovector
interaction. 

For the pseudoscalar interaction between a neutral meson field $\phi$
and a fermion field $\psi$, the Lagrangian is simply given by
\begin{equation}
{\cal L}^{\text{PS}} =-\I g\, \phi(x)\, \frac{1}{2} \bigl[ \bar{\psi}
(x), \gamma_5 \psi(x) \bigr]. \label{1}
\end{equation}
In the fifties the fermion was identified with the proton; nowadays,
$\psi$ should be associated with a quark appearing in three colors. In
our naive model the emphasis is still on electrodynamics. The only
explicit assumption of the $\psi$ and $\phi$ particles enters through
the restriction $m\gg m_\pi$. 

In order to describe the decay of the pion into two photons, we
replace the fermion fields by their vacuum expectation value in the
presence of an external electromagnetic field:

\begin{eqnarray}
{\cal L}^{\text{PS}}\to {\cal L}^{\text{PS}}_{\text{eff}} &=&
-\I g\, \phi(x)\, \frac{1}{2} \langle \bigl[ \bar{\psi}(x), \gamma_5
\psi(x) \bigr] \rangle^A \nonumber\\
&=& g\, \phi(x)\, \tr\, \gamma_5 G(x,x|A). \label{2}
\end{eqnarray}
This equation is diagrammatically represented in Fig. 1.
\begin{figure}
\begin{center}
\begin{picture}(35,35)
\put(0,5){\epsfig{figure=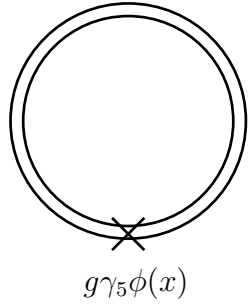,width=3.2cm}}
\put(10,0){$g\gamma_5\phi(x)$}
\end{picture}
\end{center}
\caption{Fermion loop coupled to the external electromagnetic fields
  to all orders.}
\end{figure}

The propagator $G$ satisfies the Green's function equation of a Dirac
particle
\begin{equation}
\bigl[ m+ \gamma^\mu \Pi_\mu \bigr] G(x,x'|A)=\delta(x-x'), \label{3}
\end{equation}
where $\Pi_\mu=-\I \partial_\mu -eA_\mu$. At this point we introduce
the proper-time representation for the operator $G[A]$:
\begin{equation}
G[A] =\frac{1}{m +\gamma\Pi} = \frac{m-\gamma\Pi}{m^2 -(\gamma\Pi)^2} 
=(m-\gamma\Pi) \I \int\limits_0^\infty ds\, \E^{-\I s [ m^2 -\I
  \epsilon -(\gamma\Pi)^2]} . \label{4}
\end{equation}
Inserting this representation into Eq. \re{2}, we obtain 
\begin{eqnarray}
{\cal L}^{\text{PS}}_{\text{eff}}&=& g\, \phi(x)\, \tr\, \left\{
  \gamma_5 \langle x| (m-\gamma\Pi) \I \int\limits_0^\infty ds\, 
  \E^{-\I s [ m^2  -(\gamma\Pi)^2]}|x\rangle \right\} \nonumber\\
&=&gm \, \phi(x)\,\I\,  \tr \left\{
  \gamma_5 \int\limits_0^\infty ds\, \E^{-\I sm^2} \langle x| 
  \E^{\I s(\gamma\Pi)^2}|x\rangle \right\}. \label{5}
\end{eqnarray}
Now, while identifying the loop fermion with protons, Schwinger argued
that the momentum of the outgoing photons of the pion decay is much
smaller than the mass of the proton. Therefore, the electromagnetic
fields associated with the photons vary slowly compared to the length
scale set by the Compton wavelength of the proton. As a consequence,
the constant-field approximation for the proper-time transition
amplitude appears to be appropriate in the present situation. 

So assuming that a heavy fermion (proton) runs in the loop we simply
proceed with the constant-field/low-photon-energy approximation. For
this limiting situation Eq. \re{5} yields
\begin{equation}
{\cal L}^{\text{PS}}_{\text{eff}}(x) =-\frac{1}{4} \frac{\alpha}{\pi}
\frac{g}{m} \, \phi(x)\, F_{\mu\nu} \sta{F}^{\mu\nu} =
\frac{\alpha}{\pi} \frac{g}{m}\, \phi(x)\, \mathbf{E\cdot
  B}. \label{6}
\end{equation}
This is the famous formula (5.25) in Schwinger's paper \cite{2} of
1951.

Note that although we included the coupling of the loop-fermion to all
orders to the external field, the final result \re{6} is only of
second order in the electromagnetic field strength. Hence, if we had
expanded the loop perturbatively in $\alpha$, then only the graph with
two external photons would have contributed to the final result. Note
also that we did not encounter any singular terms while calculating
$G(x,x|A)$; the dangerous terms vanished by Dirac $\gamma$-algebraic
arguments. This is not true for the pseudovector interaction which we
treat next. Its Lagrangian is given by 
\begin{equation}
{\cal L}^{\text{PV}} =-\I \frac{g}{2m} \, \partial_\mu \phi(x)\,
\frac{1}{2\I} \bigl[ \bar{\psi}(x), \gamma_5 \gamma^\mu \psi(x)
\bigr]. \label{7}
\end{equation}
Classically, this pseudovector interaction Lagrangian is formally
equivalent to the pseudoscalar counterpart as defined in Eq. \re{1},
since
\begin{eqnarray}
{\cal L}^{\text{PV}}&=&\I \frac{g}{2m}\, \phi\, \frac{1}{2\I} \Bigl\{
\bigl[ \partial_\mu \bar{\psi}, \gamma_5\gamma^\mu\psi\bigr] + \bigl[
\bar{\psi}, \gamma_5 \gamma^\mu \partial_\mu \psi\bigr] \Bigr\}
+\text{surface terms} \nonumber\\
&=& -\I g\, \phi(x) \, \frac{1}{2} \bigl[ \bar{\psi}(x), \gamma_5
\psi(x) \bigr] + \text{s.t.} \label{8}
\end{eqnarray}
In the second step, we employed the equations of motion $\gamma^\mu
\partial_\mu \psi =-\I m\psi$, $\partial_\mu \bar{\psi} \gamma^\mu =\I
m\bar{\psi}$.

However, at the quantum level, things become more
complicated. Proceeding in the same way as in the pseudoscalar case,
we naively arrive at
\begin{eqnarray}
{\cal L}^{\text{PV}}_{\text{eff}}&=&-\I \frac{g}{2m}\,
  \partial_\mu\phi(x)\, \frac{1}{2\I} \langle\bigl[ \bar{\psi}(x),
  \gamma_5\gamma^\mu\psi(x) \bigr]\rangle^{A} \nonumber\\
&=&-\I\frac{g}{2m}\, \partial_\mu\phi(x)\,
  \tr\, \gamma_5\gamma^\mu\, G(x,x|A) \nonumber\\
&\stackrel{\text{i.b.p.}}{=}&\I\frac{g}{2m}\,\phi(x)\, \partial_\mu 
  \tr\, \gamma_5\gamma^\mu\, G(x,x|A) +\text{s.t.} \label{9}
\end{eqnarray}
Now we are in trouble! Not only do we have to face the problem of
singularities in $G(x,x|A)$, but we also have to give a meaning to the
derivative at this singular coincidence point. Schwinger solved this
problem by writing
\begin{equation}
\partial_\mu  \tr\, \gamma_5\gamma^\mu\, G(x,x|A)
\to  \lim_{x'',x'\to x} 
 \Bigl\{ \bigl[ \partial_\mu' -\I e A_\mu(x')\bigr] 
    +\bigl[ \partial_\mu'' +\I e A_\mu(x'') \bigr] \Bigr\}
  \tr\, \gamma_5\gamma^\mu\, G(x',x''|A). \label{10}
\end{equation}
Now we could follow Schwinger and evaluate the right-hand side of
Eq. \re{10} in the weak-field limit, i.e., up to second order in the
field strength. This would again correspond to the triangle graph,
which is known to contribute solely to the axial-vector anomaly to any
finite order of perturbation theory.

Instead, we will try to maintain the coupling to the external field to
all orders in order to pursue the question of possible
non-perturbative contributions to the meson-photon interaction. Of
course, the price we have to pay is that we are strictly tied to the
slowly varying (constant) field approximation.

So, let us employ the representation of the fermionic Green's function
in an arbitrary constant electromagnetic field:
\begin{eqnarray}
G(x,x'|A)&=&\Phi(x,x'|A) \frac{1}{(4\pi)^2} \int\limits_0^\infty
  \frac{d s}{s^2} \left[ m-\frac{1}{2} \gamma^\mu
  [\mathsf{f}(s) +e\mathsf{F}]_{\mu\nu} (x\!-\!x')^\nu \right]
  \nonumber\\ 
&&\quad\times \exp\left[-\I m^2 s -L(s) +\frac{\I}{4}
  (x\!-\!x')\mathsf{f}(s)(x\!-\!x')\right]\, \exp\left(\I \frac{e}{2}
  \sigma Fs\right)\!,   \label{11} 
\end{eqnarray}
where
\begin{eqnarray}
\mathsf{f}(s)&=&e\mathsf{F} \, \coth (e\mathsf{F} s), \nonumber\\
L(s)&=& \frac{1}{2} \text{tr}\, \ln \frac{\sinh (e\mathsf{F}
    s)}{e\mathsf{F}s} \quad\Rightarrow\quad \E^{-L(s)} =\frac{eas\,
    ebs}{\sin eas\, \sinh ebs}, \label{12}\\
a&=& \bigl( \sqrt{{\cal F}^2+{\cal G}^2} +{\cal F}\bigr)^{1/2}, \quad 
b= \bigl( \sqrt{{\cal F}^2+{\cal G}^2} -{\cal F}\bigr)^{1/2},
\nonumber\\
{\cal F}&=&\frac{1}{4} F_{\mu\nu}F^{\mu\nu} =\frac{1}{2} \bigl(
\mathbf{B^2-E^2} \bigr), \label{13}\\
{\cal G}&=&\frac{1}{4} F_{\mu\nu}\sta{F}^{\mu\nu} =-\mathbf{E\cdot B},
\nonumber\\
\text{and}\quad\Phi(x,x'|A)&=&\exp\left[\I e \int\limits_{x'}^{x}
  d\xi_\mu  \left(A^\mu(\xi) +\frac{1}{2} F^{\mu\nu}(\xi
    -x')_\nu\right)\right]   \label{14} 
\end{eqnarray}
completely carries the gauge dependence of the propagator. Having
separated the gauge dependence in this way, we may also write
\begin{equation}
G(x',x''|A)=\Phi(x',x''|A)\, G(x',x''|A_{\text{SF}}), \label{15}
\end{equation}
where $G(x',x''|A_{\text{SF}})$ is the Green's function evaluated in
the Schwinger--Fock gauge and depends only on the field strength:
$A_{\text{SF}}^\mu=-\frac{1}{2} F^{\mu\nu}(x-x')_\nu$.  

Substituting all these results back into the starting point, i.e.,
into the effective Lagrangian in Eq. \re{9}, we find
\begin{eqnarray}
{\cal L}^{\text{PV}}_{\text{eff}}\!&=&-\frac{1}{4} 
  \frac{\alpha}{\pi}\frac{g}{m}\,\phi(x)\,\sta{F}_{\mu\nu}F^{\mu\nu}
  \label{16}\\
&&\qquad\times \lim_{x',x''\to x}\!\left\{\! \Phi(x',x''|A)
  \!\int\limits_0^\infty\! 
  d s\,  \E^{-\I m^2  s}\frac{d}{d s}
   \exp\left[\frac{\I}{4}(x'\!-\!x'')\mathsf{f}(x'\!-\!x'')\right]\!
  \right\}\!.  \nonumber
\end{eqnarray}
Comparing this with our result for the pseudoscalar interaction in
\re{6}, it is obvious that an equivalence exists between the two
different interactions on the quantum level if the limiting expression in
\re{16} finally reduces to 1 for any kind of constant
electromagnetic field. By construction, the proper-time integration
has to be performed before the limit $x',x''\to x$ can be taken. E.g.,
if we interchanged these processes in \re{16}, then we would
find a zero result, since $({d}/{d s}) \E^0=0$.

To further study the integral in Eq. \re{16} we give an explicit
expression of the function $f(s)_{\mu\nu}$:
\begin{equation}
f(s)_{\mu\nu}= \frac{1}{a^2+b^2} \bigl( a^2 \,g_{\mu\nu}
     +F^2_{\mu\nu}\bigr) eb\coth ebs
   +\frac{1}{a^2+b^2} \bigl( b^2 \,g_{\mu\nu}
     -F^2_{\mu\nu}\bigr) ea\cot eas. \label{17}
\end{equation}
For small values of $s$ we obtain
\begin{equation}
f(s)_{\mu\nu}= \frac{1}{s}\, g_{\mu\nu} +\frac{e^2}{3}\, s\,
F_{\mu\nu}^2 +{\cal O}(s^3). \label{18}
\end{equation}
and since the weak-field expansion of $f(s)_{\mu\nu}$ coincides with
the small-$s$ expansion it is here that we can make contact with
Schwinger's original calculation and so produce an effective
Lagrangian of a pseudovector interaction between a spinless meson and
a heavy fermion field in the presence of a slowly varying {\em and}
weak external electromagnetic field:
\begin{equation}
{\cal L}^{\text{PV}}_{\text{eff}} =-\frac{1}{4} \frac{\alpha}{\pi}
\frac{g}{m}\, \phi(x)\, F_{\mu\nu}\sta{F}^{\mu\nu}. \label{19}
\end{equation}
This is identical to the outcome for the pseudoscalar interaction and
constitutes the essence of Schwinger's Equivalence Theorem for the
low-energy regime. In this sense, the terminology ``low energy'' refers
to the energy of the outgoing photons (variation of the field
strength) as well as the strength of the field.

Now we could go one step further and prove the validity of the
Equivalence Theorem without the weak-field assumption. Details of the
proof can be found in our monograph \cite{9}. Our result is that
Eq. \re{19} holds for arbitrary electromagnetic field strengths as
long as the fields vary slowly compared to the Compton wavelength of
the fermionic loop particle.

It has often been emphasized in the original literature
\cite{1} that the discovery of the ABJ axial-vector
anomaly has its roots in Schwinger's work. The ABJ anomaly states that
the axial-vector current is not conserved, not only because of an
explicit breaking of the axial symmetry by a mass term, but also due
to the appearance of the $F_{\mu\nu}\sta{F}^{\mu\nu}$ term induced by
quantum vacuum effects. The celebrated result reads:
\begin{equation}
\partial_\mu\langle j_5^\mu\rangle =-2\I m \, \langle j_5\rangle +
\frac{\alpha}{2\pi}\, F_{\mu\nu} \sta{F}^{\mu\nu}. \label{20}
\end{equation}
Employing the Equivalence Theorem, we can prove Eq. \re{20}, but only
for constant external fields. Hence Schwinger's work on the
constant-field case  is only capable of deriving the anomaly in a
certain energy regime, namely, the low-energy domain. Within the usual
diagrammatic approach, Adler \cite{1} and Zumino \cite{10} arrive at
Eq. \re{20}, but this time without assuming that the electromagnetic
field has to be constant. This restriction has also been given up in
the source approach for the one-loop anomaly as presented in
Ref. \cite{3} and \cite{5}. 

Let us conclude this chapter with an interesting observation for the
constant-field case. Inserting the expression for $\langle
j_5\rangle^A$ for constant fields, 
\begin{equation}
\langle j_5\rangle^A =\I \, \tr\, \gamma_5 G(x,x|A) =-\frac{\I}{4}\,
\frac{\alpha}{\pi} \frac{1}{m} \, F_{\mu\nu} \sta{F}^{\mu\nu} \quad
\Leftrightarrow\quad {\cal L}^{\text{PS}}_{\text{eff}}
\stackrel{\re{2}}{=} -\I g\, \phi(x)\, \langle j_5\rangle^A,
\label{21}
\end{equation}
into Eq. \re{20}, we find that the divergence of the axial-vector
current vanishes: $\partial_\mu \langle j_5^\mu\rangle =0$.

This result appears to be a bit unfamiliar, because it signals the
conservation of the axial-vector current at the quantum level,
although this current is not conserved at the classical level due to
the breaking of the axial symmetry by the mass term. Therefore, the
constant-field case is an exceptional situation which creates an
``inverse anomaly'': a classically and explicitly broken symmetry is
restored by quantum effects.

Our considerations so far are strictly at the one-loop level.
Similarly to the Fujikawa \cite{11} or any other method, photonic
fluctuations have not been taken into account. This will be done in
the next chapter, in which we want to challenge the correctness of the
Adler-Bardeen theorem \cite{4}.

\section{Radiative Corrections to $\pi^0\to 2\gamma$ Decay}

Let us begin by looking again at the lowest-order triangle process. So
consider a causal arrangement in which an extended pion source emits a
pair of charged fermions that eventually annihilate to produce a pair
of photons.
\begin{figure}
\begin{center}
\begin{picture}(55,65)
\put(0,0){\epsfig{figure=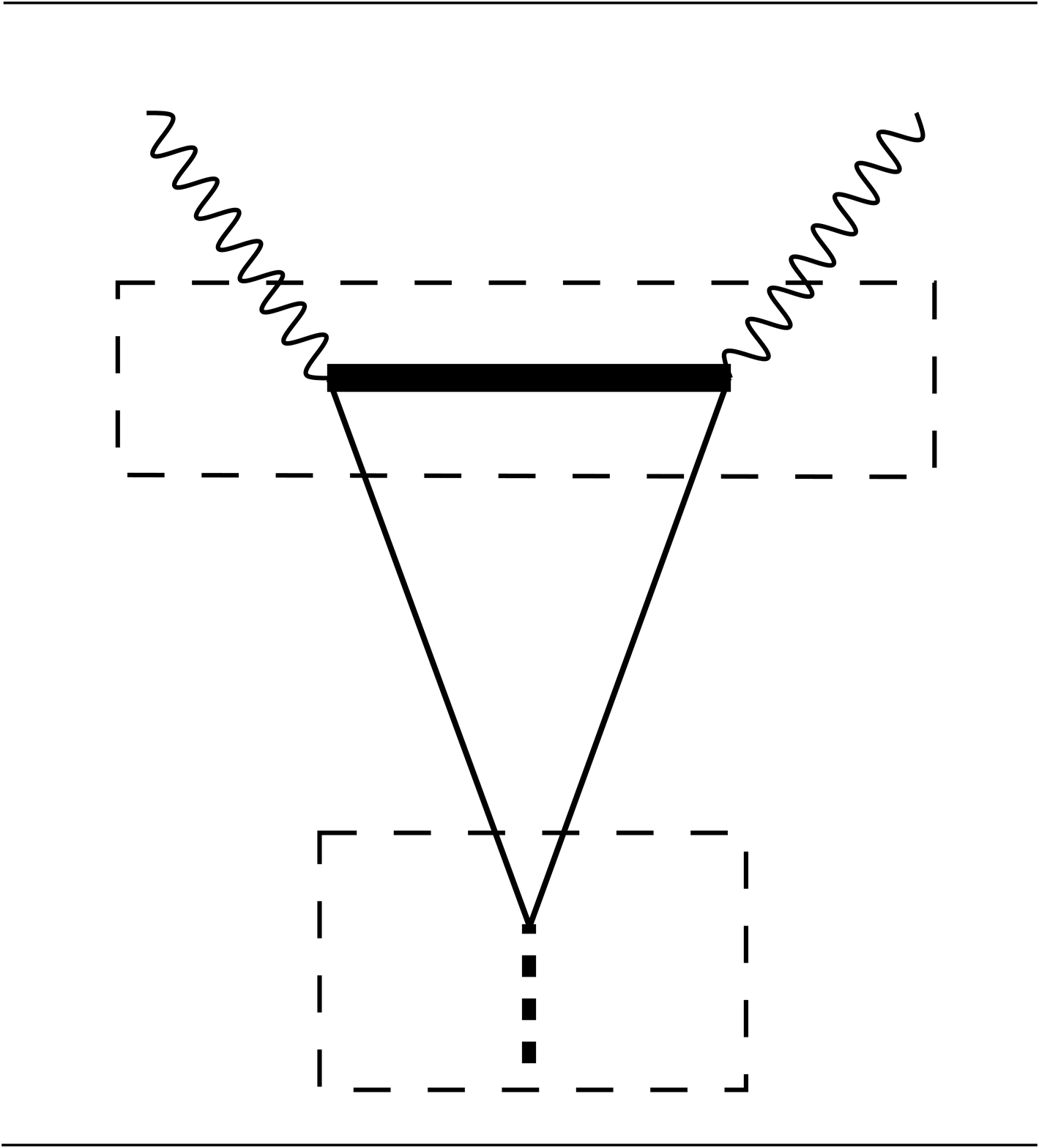,width=5.5cm}}
\put(29,7){$\pi^0$}
\put(52,3){$0_-$}
\put(52,56){$0_+$}
\put(13,52){$\gamma$}
\put(40,52){$\gamma$}
\end{picture}
\end{center}
\caption{Lowest-order causal triangle process.}
\end{figure}
The primitive interaction between the neutral pseudoscalar particle
(pion) and spin-$\frac{1}{2}$ fermions is given by
\begin{equation}
{\cal L}^{\text{PS}} =g\, \phi(x)\, \frac{1}{2} \, \psi(x)
\gamma^0\gamma_5 \psi(x), \label{22}
\end{equation} 
so that the total Lagrangian reads
\begin{displaymath}
{\cal L}=-\frac{1}{2}\, \psi \gamma^0 \bigl[ \gamma^\mu \Pi_\mu
-g\,\gamma_5\, \phi +m \bigr] \psi.
\end{displaymath}
In the present chapter we have switched to a Majorana representation
to make closer contact with the source literature.

The goal is to compute the vacuum persistence amplitude 
\begin{eqnarray}
\langle 0_+|0_-\rangle &=&\E^{\I W^{\text{PS}}_{\pi^0\to 2\gamma}}
  =\dots +\I g\int d^4x\, \phi(x) \frac{1}{2}\psi(x)
\gamma^0\gamma_5 \psi(x) +\dots \nonumber\\
\text{or}\quad W^{\text{PS}}_{\pi^0\to 2\gamma}&=& 
g\int d^4x\, \phi(x) \frac{1}{2}\psi(x) \gamma^0\gamma_5 \psi(x)
  . \label{23}
\end{eqnarray}
As for Feynman diagrams there are also standard techniques for
computing causal dia\-grams. The result for the vacuum-to-vacuum
amplitude corresponding to the causal process indicated in Fig. 2 is
given by
\begin{eqnarray}
\langle 0_+|0_-\rangle &=&\I \int d^4 x\, d^4x'\, dM^2
\frac{\alpha}{\pi} \frac{g}{m} \left( -\frac{1}{4} \right) \frac{1}{2}
\epsilon_{\kappa\lambda\mu\nu} F^{\kappa\lambda}(x) F^{\mu\nu}
\nonumber\\ 
&&\qquad \times \Delta_+(x-x';M^2)\, \phi(x')\, \frac{2m^2}{M^2} \ln
\frac{1+\sqrt{1-\frac{4m^2}{M^2}}}{1-\sqrt{1-\frac{4m^2}{M^2}}}.
\label{24}
\end{eqnarray}
This yields for the contribution to the action
\begin{equation}
W_{\pi^0\to 2\gamma}=\frac{\alpha}{\pi} \frac{g}{m} \int d^4x\,
d^4x'\, (\mathbf{E\cdot B})\, F(x-x')\, \phi(x'), \label{25}
\end{equation}
where the form factor has the momentum version
\begin{equation}
F(k^2) =\int\limits_{(2m)^2}^\infty dM^2\, \frac{2m^2}{M^2} \ln 
\frac{1+\sqrt{1-\frac{4m^2}{M^2}}}{1-\sqrt{1-\frac{4m^2}{M^2}}}\,
\frac{1}{k^2+M^2-\I \epsilon}. \label{26}
\end{equation}
It is normalized at $k^2=0$: $F(0)=1$. In the situation under
consideration $F(-m_\pi^2) =1+ \frac{1}{12} \left(\frac{m_\pi}{m}
\right)^2$, so that with $\frac{m_\pi}{m}\simeq \frac{1}{6.7}$ the
correction is about $0.2\%$. Hence for $F(x-x')\simeq \delta(x-x')$ or
$F(0)=1$ we again obtain the low-energy result corresponding to a
local effective-action term for the pion-photon coupling:
\begin{equation}
W_{\pi^0\to 2\gamma} =\frac{\alpha}{\pi} \frac{g}{m} \int d^4x (
\mathbf{E\cdot B})(x)\, \phi(x), \label{27}
\end{equation}
which is Schwinger's result from 1951 (with slowly varying fields) and
is the anomaly.

Had we treated instead of the pseudoscalar coupling the pseudovector
coupling, i.e., $g\gamma_5 \phi\to \frac{g}{2m} \I \gamma^\mu \gamma_5
\partial_\mu \phi$, our calculation would have again resulted in
expression \re{27} -- in accordance with the Equivalence Theorem.

In Eq. \re{26} we met the expression
\begin{eqnarray}
F(k^2)&=& \int\limits_{4m^2}^{\infty} \frac{dM^2}{k^2+M^2
  -\I\epsilon}\, J(M^2), \label{28}\\
\text{where}\quad J(M^2)&=& \frac{2m^2}{M^2}\, \ln  
\frac{1+\sqrt{1-\frac{4m^2}{M^2}}}{1-\sqrt{1-\frac{4m^2}{M^2}}}.
\label{29}
\end{eqnarray}
We also found that for the lowest-order result $F(k^2=0)\equiv
\tilde{I}=1$, and this is intimately related to the anomaly equation.

Now it is time to turn to the two- and three-particle exchange
processes which we put side by side with the simple triangle:
\begin{figure}
\begin{center}
\begin{picture}(140,40)
\put(0,5){\epsfig{figure=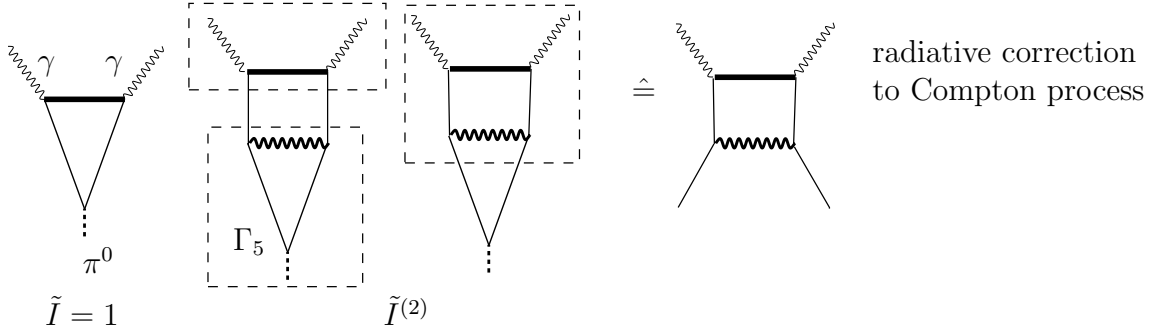,width=11cm}}
\put(10,7){$\pi^0$}
\put(4,34){$\gamma$}
\put(13,34){$\gamma$}
\put(30,10){$\Gamma_5$}
\put(5,0){$\tilde{I}=1$}
\put(50,0){$\tilde{I}^{(2)}$}
\put(115,35){radiative correction}
\put(115,30){to Compton process}
\put(83,30){$\hat{=}$}
\end{picture}
\end{center}
\caption{Two-particle exchange causal graphs.}
\end{figure}
\begin{figure}[h]
\begin{center}
\begin{picture}(140,40)
\put(0,0){\epsfig{figure=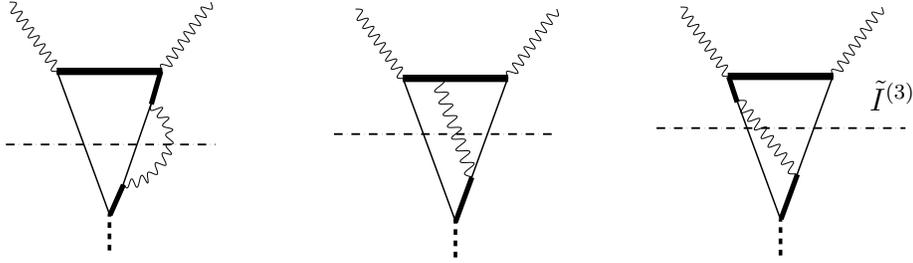,width=12cm}}
\put(115,20){$\tilde{I}^{(3)}$}
\end{picture}
\end{center}
\caption{Three-particle exchange causal graphs.}
\end{figure}
The radiative correction to the triangle process computed in
Ref. \cite{3} is obtained by adding the two- and three-particle
exchange contributions so that together with the bare triangle graph we
obtain
\begin{equation}
\tilde{I}=\tilde{I}^{(1)} +\tilde{I}^{(2)} +\tilde{I}^{(3)}
=1+\frac{\alpha}{2\pi} (1+\delta), \label{30} 
\end{equation}
where $\delta$ depends on the $\Gamma_5$ normalization point.

Miraculously, only $\delta$ is needed; all the other contributions
either cancel or yield a very simple finite expression. Hence, it is
indeed the on-shell pseudoscalar form factor that matters.

What, then, is the value of the quantity $\delta$? It must have
something to do with the normalization of the pseudoscalar form factor
$F(P^2)$. Causal analysis of diagram Fig. 5 yields
\begin{eqnarray}
F(P^2)&=& 1-\frac{\alpha}{2\pi}\, P^2\int\limits_0^1 dv\, (1+v)\,
\frac{\ln \left( \frac{4m^2}{\mu^2} \frac{v^2}{1-v^2}
  \right)}{4m^2+(1-v^2)P^2}. \label{31}\\
F(0)&=&1, \quad \mu=\text{photon mass}. \nonumber
\end{eqnarray}
\begin{figure}
\begin{center}
\begin{picture}(70,50)
\put(12,0){\epsfig{figure=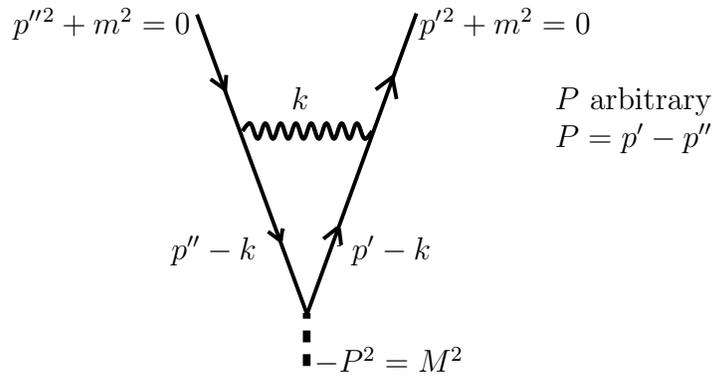,width=3cm}} 
\put(28,0){$-P^2=M^2$}
\put(60,30){$P=p'-p''$}
\put(60,35){$P$ arbitrary}
\put(25,35){$k$}
\put(-12,45){$p''{}^2+m^2=0$}
\put(9,15){$p''-k$}
\put(33,15){$p'-k$}
\put(42,45){$p'{}^2+m^2=0$}
\end{picture}
\end{center}\caption{Causal diagram for the pseudoscalar form factor.}
\end{figure}
Eq. \re{31} clearly shows that the correction to the simple triangle
anomaly is infrared sensitive. To work out the spectral weight
function that is involved in Eq. \re{31}, let us rewrite Eq. \re{31}
slightly:
\begin{eqnarray}
F(P^2)&=& 1+\alpha \int\limits_{(2m)^2}^\infty \frac{dM^2}{2\pi}
\left( -\frac{P^2}{M^2}\right)\, \frac{a(M^2)}{P^2+M^2-\I\epsilon},
\quad F(0)=1, \label{32}\\
\text{with}&&a(M^2) =\frac{(M^2-2m^2)}{M^2}
\frac{1}{\sqrt{1-4\frac{m^2}{M^2}}} \, \ln \frac{M^2-4m^2}{\mu^2}
.\label{33}
\end{eqnarray}
So far our result is expressed as a form factor multiplying the
original primitive interaction:
\begin{equation}
\langle 0_+|0_-\rangle =\I g\int d^4x\, d^4x'\, \frac{1}{2} \psi(x)
\gamma^0\, \Gamma_5(x-x')\, \psi(x) \phi(x'). \label{34}
\end{equation}
In momentum space we have
\begin{equation}
\Gamma_5(P^2) =\gamma_5 \bigl( 1+G(P^2)\bigr), \label{35}
\end{equation}
where $G(P^2)$ is given by the second term in Eq. \re{32}.
Now, since the two-particle exchange contribution (with the form
factor $\Gamma_5$ and massive photon) cannot be regularized in a
chiral invariant way, it would appear that an arbitrary normalization
point for $G(P^2)$ is allowed. Is this permitted or is there a
preferred normalization point? Let us start by introducing an
arbitrary normalization point $M_0\neq 0$ and write instead of
$-\frac{P^2}{M^2} \frac{1}{M^2+P^2}$ in Eq. \re{32} the subtracted
  form
\begin{equation}
-\frac{P^2}{M^2} \frac{1}{M^2+P^2} -\frac{M_0^2}{M^2}
 \frac{1}{M^2-M_0^2}. \label{36}
\end{equation}
Then $G(P^2)$ in Eq. \re{35} can also be written as
\begin{eqnarray}
G(P^2)&=&\alpha \int\limits_{(2m)^2}^\infty \frac{dM^2}{2\pi} \left(
  \frac{-P^2-M_0^2}{M^2-M_0^2} \right)
  \frac{a(M^2)}{M^2+P^2-\I\epsilon} \label{37}\\
\text{or}\quad G(P^2)&=& \frac{\alpha}{2\pi} \delta +\alpha
  \int\limits_{(2m)^2}^\infty \frac{dM^2}{2\pi} \left(-
  \frac{P^2}{M^2} \right) \frac{a(M^2)}{M^2+P^2-\I\epsilon},
  \label{38}\\
\text{where}\quad \delta&=& -M_0^2\int\limits_{(2m)^2}^\infty
\frac{dM^2}{M^2} \, \frac{a(M^2)}{M^2-M_0^2}. \label{39}
\end{eqnarray}

With $\mu:=\lambda m$ and $a(M^2)$ given by Eq. \re{33} we have
\begin{equation}
a(M^2)=\dots \ln \frac{M^2-4m^2}{\lambda^2 m^2} =\dots \left( \ln
  \frac{M^2-4m^2}{m^2} -2 \ln \lambda \right). \label{40}
\end{equation}
If $M_0$ were a finite number, $\delta$ would depend on $\ln \lambda$,
which is not acceptable since $\Gamma(\pi^0\to2\gamma)= 1/\tau$ would
depend on $\ln \lambda$. Hence, $M_0$ has to vanish, and then
$\delta=0$. However, if $M_0\sim \ln \frac{\mu}{m}$, one can obtain
finite $\delta$'s, e.g., $\delta=-1$, which produces no correction.
But then we would have to normalize the pseudoscalar form factor at an
infrared sensitive point. So we see that the result depends
essentially on the way the infrared regularization is performed. For
the above reason we consider the choice $\delta=0$ as the more
physically motivated regularization and this leads to the replacement
\begin{displaymath}
\frac{\alpha}{\pi} \to \frac{\alpha}{\pi} \left( 1+
  \frac{\alpha}{2\pi} \right)
\end{displaymath}
in Eq. \re{27}:
\begin{equation}
W_{\pi^0\to 2\gamma} = \frac{g}{m}\frac{\alpha}{\pi} \left( 1+
  \frac{\alpha}{2\pi} \right) \int d^4x\, \phi(x) (
\mathbf{E\cdot B})(x). \label{41}
\end{equation}
We have seen that there exist two mass scales in the theory, $\mu$ and
$\Lambda$, the two ends of a momentum flow. So when renormalizing the
theory we have to separate the renormalization constants into
infrared- and ultraviolet-sensitive parts. Both ends enter with equal
weight into the renormalization prescription. Furthermore, if two
renormalization constants have the same singular behavior at one end,
$\Lambda\to \infty$ say, then it is certainly not true that they are
equal over the whole momentum range. There are jumps on both ends of
the momentum flow of the renormalization constants whose difference
gives rise to the finite anomaly correction $\frac{\alpha}{2\pi}$.

To make contact with the work of Adler and Bardeen \cite{4} we have to
study the $\gamma_5$ vertex when the fermions are not on their mass
shell. For this reason we write down a double spectral form for the
pseudoscalar vertex function. Fig.~6 depicts the causal arrangement of
the exchange, in the presence of an external pion field, of a
fermion-photon pair between two extended fermion sources.

\begin{figure}
\begin{center}
\begin{picture}(80,60)
\put(0,0){
\epsfig{figure=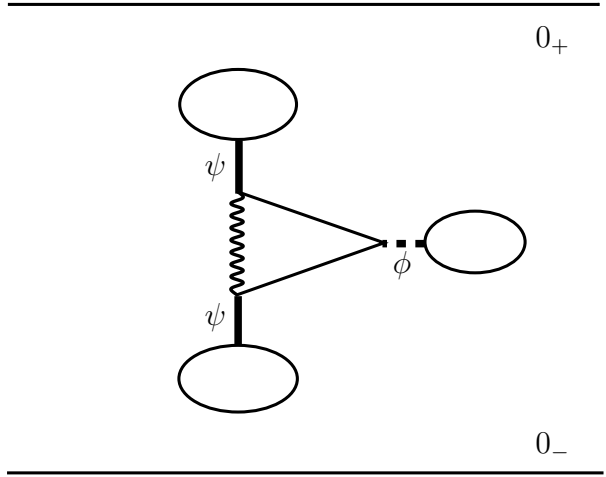,width=8cm}
} 
\put(72,3){$0_-$}
\put(72,57){$0_+$}
\put(28,20){$\psi$}
\put(28,40){$\psi$}
\put(53,27){$\phi$}
\end{picture}
\end{center}

\caption{The causal process for the pseudoscalar vertex with off-shell
  fermions.}
\end{figure}

The source-theoretical calculation yields the result
\begin{equation}
\Gamma_5(p,p')=\gamma_5 \left( 1-\frac{\alpha}{2\pi} (P^2+m\gamma P)
  \int \frac{dM^2 dM'{}^2}{\sqrt{\Delta}}
  \frac{1}{(p^2+M^2)(p'{}^2+M^2)} + \text{c.t.}\right), \label{42}
\end{equation}
where
\begin{displaymath}
\Delta=(P^2+M^2+M'{}^2)^2-4M^2M'{}^2 
\end{displaymath}
and the region of integration is bounded by
\begin{displaymath}
-\mu^2P^4+P^2[(M^2-m^2+\mu^2)(M'{}^2-m^2+\mu^2)-2\mu^2(M^2+M'{}^2)]
 \geq m^2(M^2-M'{}^2)^2. 
\end{displaymath}
The contact terms (c.t.) stand for single spectral forms plus local
functions, which have to be determined by imposing further physical
restrictions such as gauge invariance which is stated in the form of a
Ward identity:
\begin{equation}
2m\Gamma_5(p,p)=\bigl\{ \gamma_5, G_+^{-1}(p)\bigr\}. \label{43}
\end{equation}
As a further restriction to finally fix the physical $\Gamma_5$ one
finds that normalization to the on-shell result is
necessary. Altogether this brings us to an on-shell version of
$\Gamma_5$:
\begin{equation}
\tilde{\Gamma}_5(p,p')\to \gamma_5\left(1-\frac{\alpha}{2\pi}
\right)-\gamma_5 \frac{\alpha}{2\pi}\, P^2\int\limits_{4m^2}^{\infty}
\frac{dM^2}{M^2}\, \frac{a(M^2)}{M^2+P^2}, \label{44}
\end{equation}
which corresponds to $G(P^2)$ [Eq.~\re{37}] with 
\begin{equation}
\delta=-1, \quad \left( \frac{m}{M_0}\right)^2=\frac{2}{3} \ln
\frac{m}{\mu} +{\cal O}(1). \label{45}
\end{equation}
This is identical to our former physically unacceptable result which
leads to an infrared-divergent anomaly. 

If instead we choose $\delta=0$ as a physical requirement, we can
either retain the naive Ward identity \re{43} and add a constant
$\gamma_5 \gamma\cdot P$ renormalization,
\begin{equation}
\Gamma_5(p,p')=\tilde{\Gamma}_5(p,p')+\frac{\alpha}{2\pi} \gamma_5
\frac{\gamma\cdot P}{2m}, \label{46}
\end{equation}
or, alternatively, we can modify the renormalized Ward identity such
that
\begin{equation}
2m\left(1-\frac{\alpha}{2\pi}\right) \, \Gamma_5(p,p)=\bigl\{\gamma_5,
G_+^{-1}(p) \bigr\}, \label{47}
\end{equation}
which corresponds to choosing
\begin{equation}
m_0\,Z_2=\left( 1-\frac{\alpha}{2\pi}\right) m\,Z_{\text{D}},
\label{48} 
\end{equation}
where $Z_{\text{D}}$ is the pseudoscalar vertex renormalization
constant. 

\section*{Acknowledgement}
I thank H. Gies for useful discussions and carefully reading the
manuscript.

\end{document}